\documentclass[12pt]{article}

\usepackage{graphicx}
\usepackage{color}

\begin{document}
\baselineskip=20pt

\vspace{0.1cm}
\begin{center}
{\bf Physical realization and identification of topological excitations in quantum Heisenberg ferromagnet on  lattice}
\end{center}
\begin{center}
\vspace{0.5cm}
{ Ranjan Chaudhury}$^a$ and { Samir K. Paul}$^b$\\
$S.~ N.~ Bose~ National~ Centre~ For~ Basic~ Sciences , ~~ Block$-$JD,~ Sector$-$III , ~ Salt~ Lake$\\
$Calcutta$-$700098,~India$
\end{center}
\vspace{0.5cm}

a)    {$ranjan_{021258}@yahoo.com$ , ranjan@bose.res.in}

b)    { smr@bose.res.in}

\vspace{1.00cm}

Physical spin configurations corresponding to topological excitations expected to be present in the XY limit of a purely quantum spin $\frac {1}{2}$ Heisenberg ferromagnet, are probed on a two dimensional square lattice .  Quantum vortices (anti-vortices) are constructed in terms of coherent spin field components as limiting case of meronic  ( anti-meronic ) configurations . The crucial role of the associated Wess-Zumino term is highlighted in our procedure . It is shown that this term can identify a large class of vortices (anti-vortices). In particular the  excitations having odd topological charges form  this class and  also  exhibit a self-similar pattern regarding the internal charge distribution .This manifestation of different behaviour of the odd and the even topological sectors is very prominent in the strongly quantum regime but fades away as we go to higher spins. Our formalism is distinctly different from the conventional approach for the  construction of quantum vortices ( anti-vortices ).

\vspace{0.2cm}

{\bf PACS}     : 75.10. Jm ; 03.70. +k ; 03.75. Lm\\
{\bf Key~Words} : Topological, Ferromagnet, Wess Zumino, Vortices.
\vspace{10.00cm}

{\bf 1. Introduction}\\

\vspace{0.1cm}
The physical existence of topological excitations in quantum  spin systems in one and two dimensions can be investigated by making use of various field theoretic techniques such as  quantum action-angle  representation, coherent state formulation and field theoretic approach manifesting a 'Berry Phase'  [1-3]. The XY-limit of  Heisenberg ferromagnet  also belongs to this class of systems .  It has been shown that in the case of a Heisenberg antiferromagnetic chain, the Wess-Zumino ( WZ) term in the effective action  is genuinely a topological term [1]. Moreover in the long wavelength limit of this system the explicit expression of this term is similar to that of the topological term for a nonlinear sigma model. Extending this approach further, it was shown by us [3]  that it is possible to express the WZ term as a topological term, also in the case of a ferromagnetic chain in the long wavelength limit. Later we generalized this to the   case of two dimensional ferromagnet [3,4] where again we demonstrated that in the above limit we could indeed get an expression from the WZ term , indicating the possibility of topological excitations.\\  
 
 Thus at this stage of analysis it is necessary to examine the physical configurations of these excitations in terms of coherent fields as well as the consequence of the WZ term, explicitly on a two dimensional square lattice. At the moment we do this for the XY-limit of the two dimensional Heisenberg ferromagnet with spin $\frac {1}{2}$ . We show that the WZ term can indeed identify a large class of the  topological excitations . Furthermore we demonstrate that this term can clearly  differentiate between vortices ( and antivortices ) with different charges within this class. It is important to note that our entire analysis is valid for all temperatures . \\
{\bf 2~Mathematical Formulation}\\
2.1 Action and the Wess-Zumino term\\
  The quantum Euclidean action ${{\mathcal S}_E}$ for the spin coherent fields 
${\bf n}(t)$ , on a single lattice site , can be written as [2,3] \\
\begin{equation}
{{\mathcal S}_E} = -is{{\mathcal S}_{WZ}}[{\bf m}] + {\int_0^{\beta}}dt {\mathcal H}({\bf n})     
\end{equation}       
where $s$ is the magnitude of the spin ($s = {\frac{1}{2}}$ in the present case ) and 
\begin{eqnarray}
\langle {\bf n}\vert {\bf S}\vert {\bf n}\rangle & = & s{\bf n} \nonumber\\ 
{\mathcal H}({\bf n}) & = &  \langle {\bf n}\vert {\mathcal H}({\bf S})\vert {\bf n}\rangle 
\end{eqnarray}
${\mathcal H}({\bf S})$ being the single site effective spin Hamiltonian in spin ($s$) representation  and ${\mathcal H}({\bf n})$ is the corresponding Hamiltonian in the coherent spin fields , $\bf n(t)$ . The variable $t$ denotes the pseudo-time having the dimension of inverse temperature with $\beta = {\frac{1}{kT}}$ as usual , $T$ being the real thermodynamic temperature of the spin system.\\
 The Wess-Zumino term ${{\mathcal S}_{WZ}}$ is given by (see Fradkin in Ref.[2])
\begin{equation}
{{\mathcal S}_{WZ}}[{\bf m}]={\int_0^{\beta}}dt{\int_0^1}d{\tau} {\bf m}(t,{\tau})\cdot {\partial_t}{\bf m}(t,{\tau})\wedge {\partial_{\tau}}{\bf m}(t,{\tau})
\end{equation}
with ${\bf m}(t,0)\equiv {\bf n}(t)$, ${\bf m}(t,1)\equiv {{\bf n}_0}$, and
${\bf m}(0,{\tau})\equiv {\bf m}(\beta ,\tau )$, $t\in [0,\beta ]$, $\tau \in [0,1]$ .   \\
 The geometrical interpretation  of equation(3) is that the left hand side represents the area of the cap bounded by the trajectory  $\Gamma$ , parametrized by ${\bf n}(t)$ $[{\equiv} ({n_1}(t),{n_2}(t),{n_3}(t))]$  on the sphere:   
\begin{equation}
{\bf n}(t)\cdot {\bf n}(t) = 1
\end{equation}
Furthermore the fields ${\bf m}(t,{\tau})$ are the fields in the higher dimensional $(t, \tau)$-space and the boundary values ${\bf n}(t)$ are the coherent spin fields . ${{\bf n}_0}$  is the fixed point $(0,0,1)$ on the sphere and the ket $\vert {\bf n}(t) \rangle$ appearing in the right hand side of equation (2), is the spin coherent state on a single lattice point [1-3] .For simplicity we have suppressed the coordinates in all the expressions .\\
  For the whole two dimensional lattice , we express ${\bf n}(ia ,ja , t)$ as   ${\bf n}(ia ,ja )$ for breavity , $a$ being the lattice spacing . The WZ term for the whole lattice can formally be written as 
\begin{equation}
{{\mathcal S}_{WZ}^{tot}}  = {\sum_{i,j=1}^{2N}} {{\mathcal S}_{WZ}}[{\bf m}(ia,ja)]
\end{equation}
However it is  possible to evaluate only the 'difference' of ${{\mathcal S}_{WZ}}[{\bf m}(ia,ja)]$ terms at two neighbouring lattice sites in terms of the coherent spin fields ${\bf n}(ia,ja)$. Thus to extract the topological-like contribution from the total $WZ$ term ${{\mathcal S}_{WZ}^{tot}}$, we use the following expressions for the above 'difference' (see Fradkin in Ref.[2]) :
\begin{eqnarray}
{\delta_x}{{\mathcal S}_{WZ}}[{\bf m}(\bf r)] & = & {{\mathcal S}_{WZ}}[{\bf m}(ia,ja)] - {{\mathcal S}_{WZ}}[{\bf m}((i-1)a,ja)]\nonumber\\
& = &{\int_0^{\beta}}dt [{\delta_x} {\bf n}\cdot ({\bf n}\wedge {\partial_t}{\bf n})]({\bf r})\nonumber\\
{\delta_y}{{\mathcal S}_{WZ}}[{\bf m}(\bf r)] & = & {{\mathcal S}_{WZ}}[{\bf m}(ia,ja)] - {{\mathcal S}_{WZ}}[{\bf m}(ia,(j-1)a)]\nonumber\\
 & = &{\int_0^{\beta}}dt[{\delta_y} {\bf n}\cdot ({\bf n}\wedge {\partial_t}{\bf n})]({\bf r})
\end{eqnarray}
where
\begin{eqnarray}
{\delta_x}{\bf n}(ia,ja) & = &{\bf n}( ia,ja ) - {\bf n}( (i-1)a,ja )\nonumber\\{\delta_y}{\bf n}(ia,ja) & = &{\bf n}( ia,ja ) - {\bf n}( ia,(j-1)a )
\end{eqnarray}
 with ${\bf r}\equiv (ia,ja)$ ,  and $i,j = 1,2,.....,2N$.\\
We now write the following identity for the right hand side of equation (5):
\begin{eqnarray}
{{\mathcal S}_{WZ}^{tot}} & = &{\sum_{i,j=1}^{2N}} {{\mathcal S}_{WZ}}[{\bf m}(ia,ja)]\nonumber\\
& = &{\sum_{i,j=1}^{2N}} [{\frac{1}{2}}{{\mathcal S}_{WZ}}[{\bf m}((i-1)a,ja)] + {\frac{1}{2}}{{\mathcal S}_{WZ}}[{\bf m}(ia,(j-1)a)]\nonumber\\ 
&   & +{\frac{1}{2}}{\delta_x}{{\mathcal S}_{WZ}}[{\bf m}(ia,ja)] +{\frac{1}{2}}{\delta_y}{{\mathcal S}_{WZ}}[{\bf m}(ia,ja)]] 
\end{eqnarray}
The philosophy behind the above exercise was to extract the quantity which is directly calculable viz. difference in $WZ$ term and cast it in the form analogous  to that of the winding number in the case of a one dimensional system [2].\\
From equations (6), (7) and (8) we get
\begin{eqnarray}
2{{\mathcal S}_{WZ}^{tot}} & = & {\sum_{i,j=1}^{2N}} \{ {{\mathcal S}_{WZ}}[{\bf m}( (i-1)a, ja )] + {{\mathcal S}_{WZ}}[{\bf m}( ia, (j-1)a )]          \nonumber\\
&   & -{\int_0^{\beta}}dt {\bf n}( (i-1)a,ja )\cdot ({\bf n}\wedge {\partial_t}{\bf n})( ia,ja )\nonumber\\
&   &-{\int_0^{\beta}}dt  {\bf n}( ia,(j-1)a )\cdot ({\bf n}\wedge {\partial_t}{\bf n})(ia,ja) \}
\end{eqnarray}
The pseudo-time derivatives of the coherent spin fields, $\partial_t{\bf n}(ia,ja)$ in the above equation can be expressed in terms of coherent fields at the above site and at the nearest neighbour sites through equations of motion which can be obtained from the action corresponding to the appropriate spin model on the lattice   (see $Eqns. (13),~~(14)$ and $Eqns. (A.1),~~(A.2),~~(A.3)$ in the appendix) . This brings out the topological character of the last two terms in the above expression of ${{\mathcal S}_{WZ}^{tot}}$. However the remaing terms (viz., 1st and the 2nd) in equation (9) have no such properties and hence we do not consider them for determination of topological charge . Therefore we keep only the topologically relevant terms and construct the following effective $WZ$-term :
\begin{eqnarray}
2{{\mathcal S}_{WZ}^{effective}} & = & - {\sum_{i,j=1}^{2N}} \{ {\int_0^{\beta}}dt {\bf n}( (i-1)a,ja )\cdot ({\bf n}\wedge {\partial_t}{\bf n})( ia,ja )\nonumber\\
&   & + {\int_0^{\beta}}dt  {\bf n}( ia,(j-1)a )\cdot ({\bf n}\wedge {\partial_t}{\bf n})(ia,ja) \}
\end{eqnarray}

2.2  {\bf Action corresponding to spin model and equations of motion}\\
  The spin Hamiltonian corresponding to nearest neighbour anisotropic Heisenberg ferromagnet of $XXZ$ type is given by[3]
\begin{equation}
{\mathcal H}({\bf S}) = - g{\sum_{\langle {\bf r},{{\bf r}^{\prime}}\rangle}}{\bf {\tilde S}}({\bf r})\cdot {\bf {\tilde S}}({\bf {r^{\prime}}}) - g{\lambda_z}{\sum_{\langle {\bf r},{\bf {r^{\prime}}}\rangle}}{S_z}({\bf r}){S_z}({\bf {r^{\prime}}})
\end{equation} 
with $g\ge 0$ and $0\le {\lambda_z}\le 1$ , $\bf r$,$\bf {r^{\prime}}$ running over the lattice, and $\langle {\bf r},{\bf {r^{\prime}}}\rangle$ signifies nearest neighbours with ${\bf S}\equiv ({\bf {\tilde S}}, S_z)$.\\
It follows from equations $(2)$ and $(11)$ that the total spin Hamiltonian on the lattice , in terms of coherent fields is given by 
\begin{equation}
{\mathcal H}({\bf n}) = -g{s^2} [ {\sum_{\langle (i,j),({i^{\prime}} ,{j^{\prime}})\rangle}}{\bf {\tilde n}}(ia,ja)\cdot {\bf {\tilde n}}({i^{\prime}}a ,{j^{\prime}}a) + {\lambda_z}{\sum_{\langle (i,j),({i^{\prime}} ,{j^{\prime}})\rangle}}{n_z}(ia,ja) {n_z}({i^{\prime}}a ,{j^{\prime}}a) ]   
\end{equation}
Using equation (1) we can write the quantum action (Euclidean)for the two dimensional anisotropic ferromagnet as:
\begin{eqnarray}
{{\mathcal S}_E^{lattice}}& = &-is{\sum_{i,j}}{{\mathcal S}_{WZ}}[{\bf m}(ia,ja)] + {\int_0^{\beta}}dt [-g{s^2}{\sum_{\langle (i,j),({i^{\prime}} ,{j^{\prime}})\rangle}}{\bf {\tilde n}}(ia,ja)\cdot{\bf {\tilde n}}({i^{\prime}}a ,{j^{\prime}}a)\nonumber\\
&  & - g{\lambda_z}{s^2}{\sum_{\langle (i,j),({i^{\prime}} ,{j^{\prime}}) \rangle}}{n_z}(ia,ja){n_z}({i^{\prime}}a ,{j^{\prime}}a)]
\end{eqnarray}
where $s=\frac {1}{2}$, since we are considering the extreme quantum case.\\
The above action ${{\mathcal S}_E^{lattice}}$ is used to derive equations of motion as described in the previous section. We minimise ${{\mathcal S}_E^{lattice}}$ subject to the local constraint (4) on each lattice site. Thus we minimise the following quantity:\\
\begin{equation}
{{\mathcal F}_{tot}} = {{\mathcal S}_E^{lattice}} +  {\frac{1}{2}}{\int_0^{\beta}}~dt~{\sum_{(i,j)}}~{a^2}{\lambda_{i,j}} [~{{\bf n}^2}(i,j) - 1] \}            \end{equation}
The 2nd term in the right hand side of the above equation is the lattice version of the term ${\int} {d^2}x {\int_0^{\beta}} {\lambda} ({\bf x}, t)({{\bf n}^2}({\bf x},t) - 1)$ where ${\lambda}({\bf x},t)$  is an auxiliary field playing the role of a multiplier . The equations of motion which follow from the above minimisation are given in the Appendix $A$ . We substitute the expressions for ${\partial_t}{\bf n}$ from $Eqns.$ ( $A.1$ ) into $Eqn. (10)$.  The purpose of this exercise  is to obtain an approximate expression for  ${{\mathcal S}_{WZ}^{effective}}$ given by equation (10) corresponding to the $XXZ$ Heisenberg ferromagnet.

\vspace{0.5cm}

3. {\bf Calculations and Results}\\
Now we present calculations for ${{\mathcal S}_{WZ}^{effective}}$ corresponding to vortices (anti-vortices) constructed out of coherent spin field components 
${n_1}(ia , ja)$, ${n_2}(ia , ja)$ and  ${n_3}(ia , ja)$ . We analyse all the higher charged vortices by decomposing them in terms of elementary plaquetts of unit charge . We find that the odd-charged  vortices can be described consistently by  a  flattened  meron configuration . Even-charged vortices however , behave
in an anomolous way . 
 
3.1 {\bf Analysis of 1-vortex}\\
\begin{figure}[!htbp]
\begin{center}
\includegraphics[keepaspectratio,width=7cm]{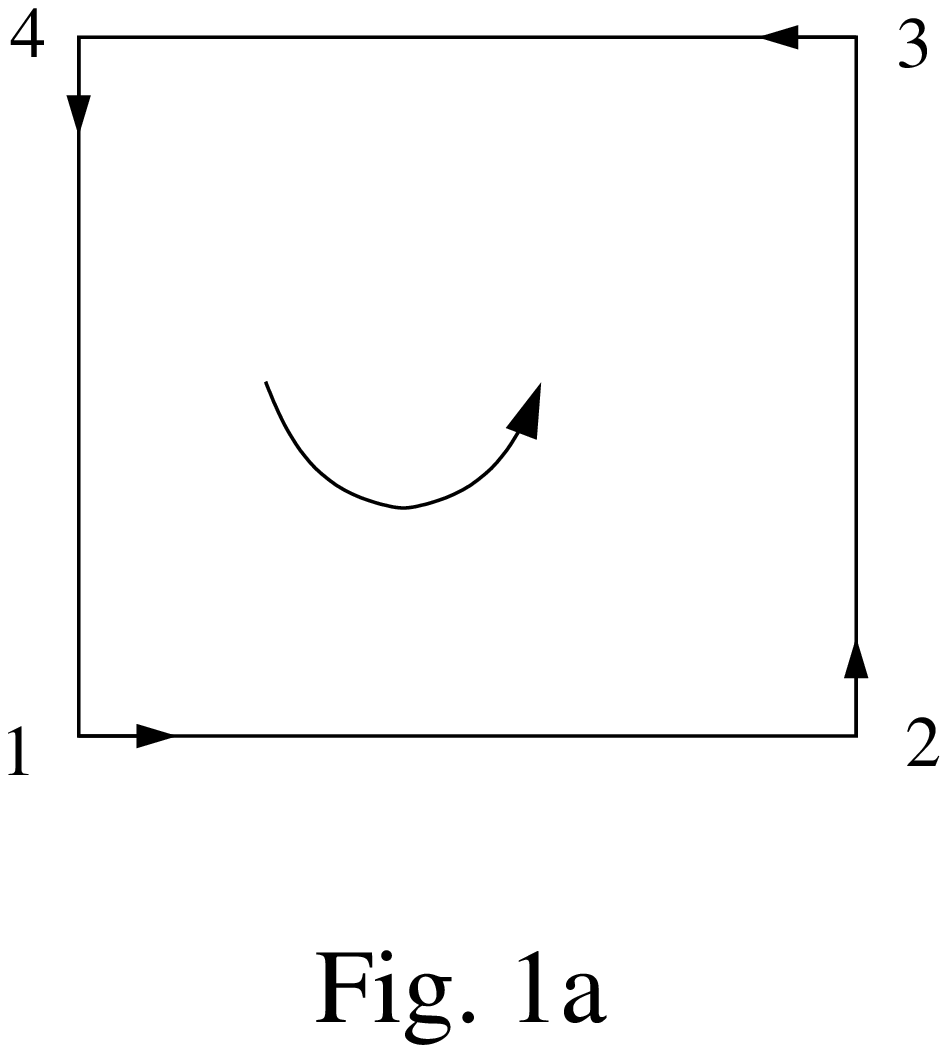}
\includegraphics[keepaspectratio,width=7cm]{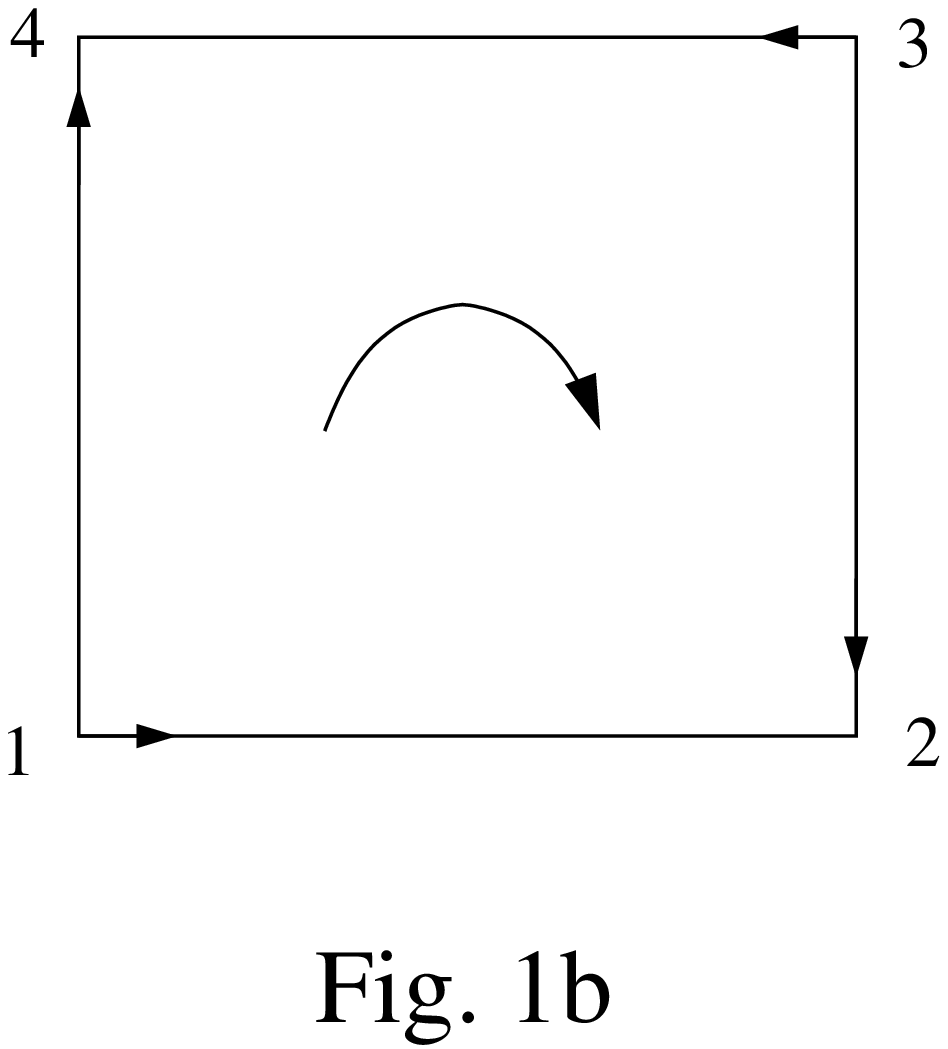}
\end{center}
\caption{(a) 1-vortex , (b) 1-anti-vortex}
\end{figure}
Equation (10) shows that the summands are associated with the  vertices   2 , 3  and 4 where the vertices 1, 2, 3 and 4 of the 1-vortex [$see~Fig.1$]   are assigned  the coordinates  $(i-1)a , (j-1)a$,    $(ia,(j-1)a)$ , $(ia,ja)$  and $((i-1)a,ja)$ respectively .  The contribution  of ${{\mathcal S}_{WZ}^{effective}}$ to a single plaquette [$see~Fig.1$] with  the above vertices can be written as
(see Appendix) :
\begin{eqnarray}
{{[{{\mathcal S}_{WZ}^{effective}}]}_{(1-vortex)}} & = & -  \{ {\int_0^{\beta}}dt [{\bf n}((i-1)a,ja)\cdot ({\bf n}\wedge {\partial_t}{\bf n}) (ia,ja)\nonumber\\
&  & +{\bf n}(ia,(j-1)a)\cdot ({\bf n}\wedge {\partial_t}{\bf n}) (ia,ja)\nonumber\\
&  & + {\bf n}((i-1)a,(j-1)a)\cdot ({\bf n}\wedge {\partial_t}{\bf n}) (ia,(j-1)a)\nonumber\\
&  &+{\bf n}((i-1)a,(j-1)a)\cdot ({\bf n}\wedge {\partial_t}{\bf n}) ((i-1)a,ja)
] \}
\end{eqnarray}                  
Now we focus our attention to the physical construction of vortices corresponding to our specific model . As we are interested in the extreme $XY$- anisotropic limit, we  assume ${n_3} ({i^{\prime}}a ,{j^{\prime}}a) =  \sin \epsilon ({i^{\prime}}a ,{j^{\prime}}a)$ at each lattice point $({i^{\prime}}a ,{j^{\prime}}a)$     where $\epsilon ({i^{\prime}}a ,{j^{\prime}}a) $ is a very small positive quantity for each $({i^{\prime}} ,{j^{\prime}})$. Then it follows from  $Eqn.(4)$  that     within a vortex (or anti-vortex) the following conditions must be satisfied :\\    
If one of the components ${n_1}$ (or ${n_2}$)  at a vertex $(la,ma)$ for $(l,m)$ =$((i-1) , (j-1))$,    $(i,(j-1))$ , $(i,j)$ and $((i-1),j)$ is given by $\pm (1-\delta)$ then  the other one is given by $\pm {\sqrt {[{{\cos}^2}{\epsilon (la,ma)}  - {{(1-\delta )}^2}]}}$ ; $\delta$ being small and positive , and assuming the same value at  each vertex  of  the  plaquette .\\  
For illustration (see Fig.-1) we have a quantum vortex of charge $+1$ in which the horizontal arrow $\rightarrow$ at a vertex represents  $n_1$ with value $1-\delta$ and the vertical arrow $\uparrow$ implies $n_2$ with value $1-\delta$. Further, in this figure the horizontal arrow $\leftarrow$ at a vertex denotes $n_1$ having value  $-({1-\delta})$ and the vertical arrow $\downarrow$ represents $n_2$ with value $-({1-\delta})$. As shwon in the appendix ; since the horizontal or vertical arrow has value  $\pm ({1-\delta})$ for all time ,the quantity  $\delta$ must be independent of time . It may be remarked that the vortex configuration (shown in Fig.1a) has a topological charge $+1$ , as the spin rotates through an angle $+2\pi$ in traversing the boundary once in the anti-clockwise sense.\\
 In this connection let us point out that usually in a two dimensional vortex  corresponding to spin $\frac {1}{2}$ quantum spin model, the arrows $\rightarrow$ and  $\uparrow$ signify eigenstates of $S_x$ and $S_y$ respectively having  eigenvalues $+{\frac{1}{2}}$  [6]. The states $\leftarrow$  and $\downarrow$ are eigenstates of these operators respectively having eigenvalues $-{\frac{1}{2}}$ . In our present formulation for quantum vortices, we make use of  the spin coherent field components $n_1$ and  $n_2$ to form vortex (anti-vortex), with ${n_3}(ia , ja) = \sin \epsilon (ia,ja)$  at each lattice point. Our picture is approximately that of  flattened "meron(anti-meron) configurations" [7,8] mimicking a  vortex (or anti-vortex) . \\
  This configuration arises naturally only when $\lambda_z \longrightarrow  0$ $i.e.$ in a pure XY model . As explained before, in this quantum vortex configuration  $n_1$ or  $n_2$ cannot be exactly equal to $\pm 1$ . Rather we will have to choose them as $\pm (1-\delta )$, where $\delta$ is an infinitesimal quantity.  More detailed  reasons are as follows:\\
 If we want to identify the spin state $\vert \rightarrow \rangle$ (eigenstate of the operator $S_x$ having eigenvalue $+ {\frac{1}{2}}$) with the coherent state $\vert {\bf n} \rangle = {\cos{\frac{\theta}{2}}}{\vert  {\frac{1}{2}}\rangle} + ({e^{-\phi}}) {\sin{\frac{\theta}{2}}}{\vert  {-\frac{1}{2}}\rangle}$ corresponding to $s = {\frac {1}{2}}$, at a vertex $[see Fig.1]$, we  have ${n_1}=1,{n_2}=0~and~{n_3}=0$ at that vertex. Similarly for $\vert \leftarrow \rangle$ we put ${n_1}=-1,{n_2}=0~and~{n_3}=0$ ; for $\vert \uparrow \rangle$ we  have ${n_1}=0,{n_2}=1~and~{n_3}=0$ and to represent   $\vert \downarrow \rangle$   we require ${n_1}=0,{n_2}=-1~and~{n_3}=0$ . In this way we can represent spin states at each vertex by the corresponding  coherent spin states [see $Eqn.(2)$]. These assignments however violate the non-zero magnititude of  ${n_3}$ , which is given by ${n_3} (la,ma)= \sin{\epsilon} (la.ma)$ at a vertex point of the vortex plaquette. Therefore we must choose ${n_1}$ or ${n_2}$ to be $\pm (1-\delta )$ to preserve the constraint given by $Eqn.(4)$    .\\
Let us now look at the symmetry properties of ${{\mathcal S}_{WZ}^{effective}}$ . It can be shown that (see appendix) ${{\mathcal S}_{WZ}^{effective}}$ can easily be decomposed into two parts such that the first part changes sign in going over from a 1-vortex configuration to the corresponding anti-vortex configuration ; whereas the second part remains invariant under this operation. We denote the first part by $B$ and the second part by $A$. This transformation from vortex to anti-vortex can be implemented by changing ${n_2}(2)$ and ${n_2}(4)$ in Fig.1 to -${n_2}(2)$ and -${n_2}(4)$ respectively.
Algebraically this means that $A$ contains terms which are  of even degree in ${n_2}(2)$ and ${n_2}(4)$ whereas $B$ contains terms that are linear or of odd degree in ${n_2}(2)$ and ${n_2}(4)$. From $Eqn.(15)$ we have ${{\mathcal S}_{WZ}^{effective}}$  in the form $ A + B$ by using $Eqn.(15)$  as explained in the Appendix A. :\\

3.2 {\bf Analysis of 2-vortex} \\
 For a typical 2-vortex we refer to $Fig.2$ . We  calculate the contribution of  ${{\mathcal S}_{WZ}^{effective}}$ given by equation (10) on such a plaquette by algebraically adding the contributions of ${{\mathcal S}_{WZ}^{effective}}$ on  each of the individual elementary plaquettes (subvortices) with a weightage factor of $\frac{1}{2}$ to the common bonds shared between the pairs of adjacent subvortices.  We denote these subvortices by $a$ , $b$ , $c$ and $d$, each carrying topological charge $+1$ [ $Fig.2$ ] .  We are interested in those field configurations for which the  contributions  of ${{\mathcal S}_{WZ}^{effective}}$  on the common bonds cancel each other (see Appendix) and only the peripheral contribution on the boundary remains. However the spin at the central lattice point of the vortex ( the point O in $Fig.2$ ) becomes ambiguous , as is clear from the fact that the spin configurations at the lattice points on the boundary of the plaquette cannot be shrinked to a unique spin at the centre . This is reflected through the fact that the horizontal effective spin at the centre vanishes. Thus the central point turns out to be a $\bf 'defect'$ or a singular point. This brings pecularities in the behaviour of the neighbouring spins as well ,  leading to the  breakdown  of the 2-vortex in the flattened meron configuration limit, if we construct the vortices in terms of elementary plaquetts . It can be explicitly demonstrated that this scenerio persists in the case of all other even-charged vortices as well ( see Appendix )   \\

  Thus the contributions of $ {{\mathcal S}_{WZ}^{effective}}$  on the boundary of the 2-vortex is not well defined . The scenerio persists in all the vortices (anti-vortices)  possessing $\bf even~~valued$ topological  charges , as can be read out from the spin field configuration in the case of 4-vortex $[see~~Fig.4]$  . \\     
\begin{figure}[!htbp]
\begin{center}
\includegraphics[keepaspectratio,width=10cm]{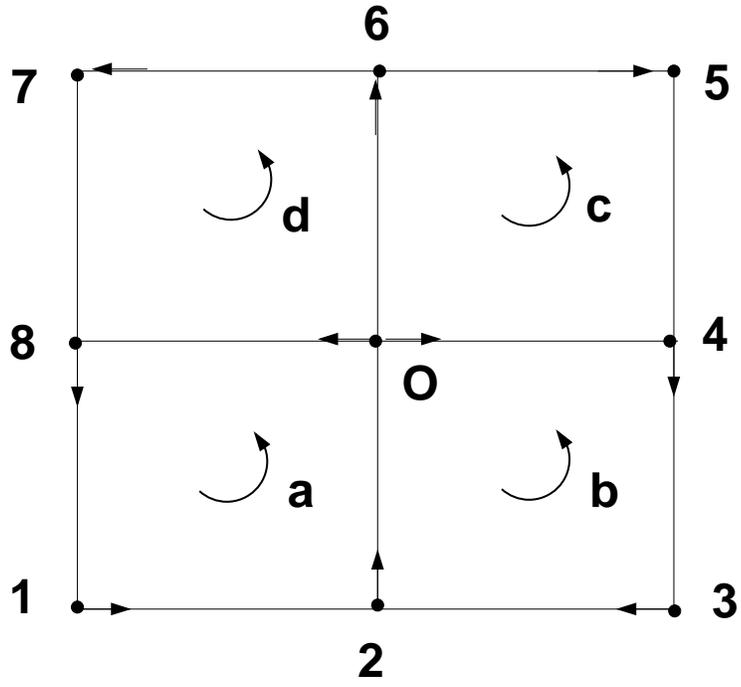}
\end{center}
\caption{2-vortex}
\end{figure}

\vspace{0.2cm}

3.3 {\bf Analysis of 3-vortex} \\
\begin{figure}[!htbp]
\begin{center}
\includegraphics[keepaspectratio,width=10cm]{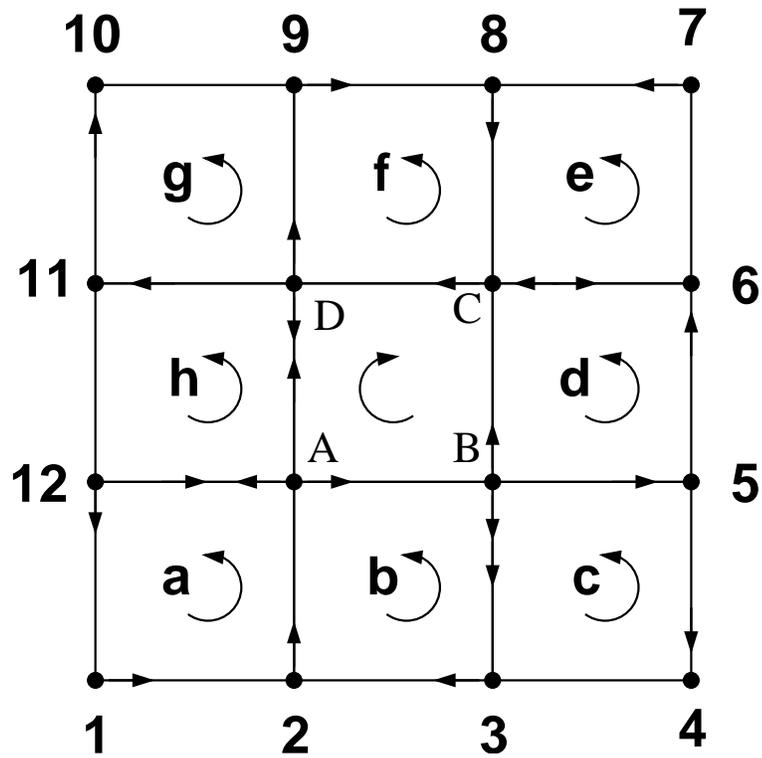}
\caption{3-vortex}
\end{center}
\end{figure}
 For 3-vortex [see Fig. 3] however , we have a consistent spin field configuration . We have now an elementary anti-vortex plaquette at the central region with well defined field configurations , the contributions of ~${[{{\mathcal S}_{WZ}^{effective}}]}$~ along the common bond cancel  giving rise to  consistent spin field configurations .
 It may be pointed out that in contrast to the even-valued charge case we now have a subvortex with opposite charge occupying  the central region. Studying the configuration in  $Fig.~3$ , we discover the following identity for a topological Q-vortex ,with odd values of Q ( $Q \ne 1$ ), which describes the  topological charge  distribution inside the vortex  consistently.
\begin{equation} 
Q = {\frac{1}{4}} [{Q^2} - {{(Q - 2)}^2}] + 1
\end{equation}
The quantity $(Q-2)$ is the absolute value of the effective charge ( negative ) of the core which is an anti-vortex in this case. This can be generalized for Q-antivortex also. It may be pointed out  that this equation is obtained directly by studying the charge distribution within these configurations . Notice that the above equation holds good even within the core, i.e., when $Q$ is replaced by $(Q-2)$, and thereby depicts a ${\bf self~similar~pattern}$.

3.4 {\bf Analysis of 4-vortex} \\
\begin{figure}[!htbp]
\begin{center}
\includegraphics[keepaspectratio,width=10cm]{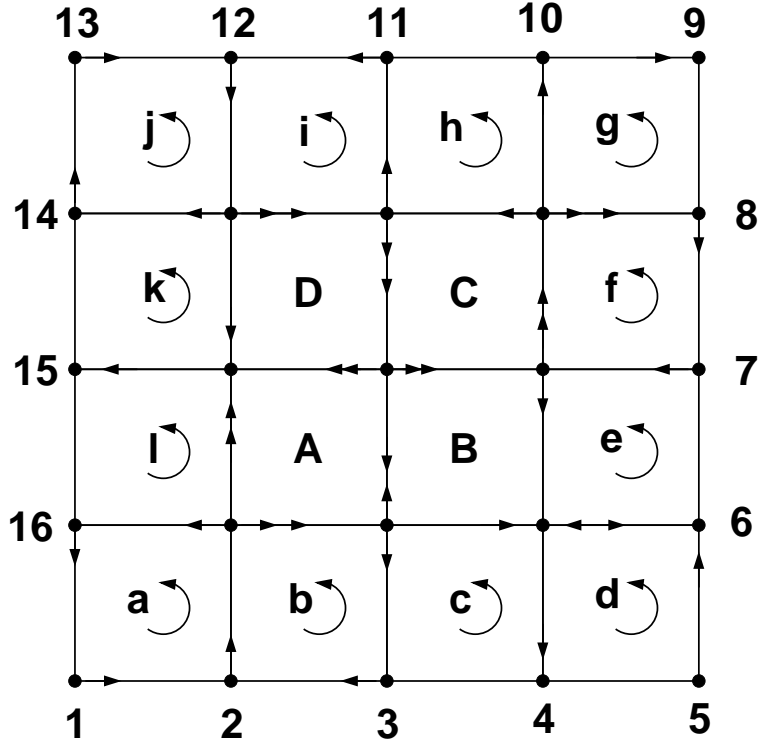}
\end{center}
\caption{4-vortex}
\end{figure}
Here  again  [see $Fig.~4$]  the spin at the centre becomes ambiguous,  the horizontal effective spin at the centre vanishes . Once again the central point  turns out to be a defect or a singular point, rendering cancellation  of the  contributions  of ${{\mathcal S}_{WZ}^{effective}}$  on the common bonds impossible .The  ambiguity of the suitable  spin field configuration  for 4-vortex is explained through equation (~~) in the appendix.

{\bf 4. Conclusion and Discussion}\\
i) Our calculations and analysis with  spin $\frac {1}{2}$ ferromagnetic quantum $XY$ model in two dimensions clearly bring out distinguishing features between  even and odd charge sectors. The internal charge distribution equation [equation (30)] is obeyed by the excitations with odd charge in totality. The even charge sectors however are not governed by this equation.\\
  
ii) Our work has established the role of WZ term as a 'topological charge measuring quantity' , obtained from microscopic theory. Furthermore, this term is able to test various internal consistencies and conservation conditions involving the topological charge distribution for a large class of  excitations . Thus our approach is more powerful than that based on the heuristic operators  suggested for determining the  charges of topological excitations [6,9].\\
 
iii) Our future plan includes the generalization of our approach to the case of finite $\lambda_z$ to achieve a physical realization  of excitations  of meronic type in the spin models . Furthermore , making use of these results and these configurations  we aim to evaluate the static and dynamic spin correlations for spin $\frac {1}{2}$ anisotropic quantum Heisenberg ferromagnet at any temperature  in two dimensions . These will contain contributions from both merons and  anti-merons and are expected to exihibit distinguishing features from both even and odd topological charge sectors . These are  of great importance in  analysing the results from inelastic neutron scattering experiments, carrying the signature of the dynamics of these configurations themselves [7] . Our procedure can also throw some light on the possible phenomenological scenario of quantum Kosterlitz - Thouless transition [3,6,10].\\

iv) Last but not the least, the present investigation and methodology of ours may further be extended to low dimensional fermionic models[11] as well.\\

 To conclude, this study of topological spin excitations on the 2D-lattice will undoubtedly play an important role in the understanding of thermodynamics of low dimensional ferromagnets.\\ 
{\bf Acknowledgement} : One of the authors (SKP) would like to thank M.G.Mustafa, Rajarshi Roy and Purnendu Chakraborti for their valuable help in the preparation of the figures.

\vspace{0.1cm}

{\bf References}\\
$[1]$  Fradkin E and  Stone M 1988 Phys. Rev. B 38  7215 ; Haldane F D M 1983 Phys. Rev. Letts. $\bf 50$ 1153 ; Mikeska H J 1978 Jour. of. Phys. C $\bf 11$  L29 \\
$[2]$  Fradkin E 1991 Field Theories of Condensed Matter Systems ( Addision-Wesley     , CA) , Haldane F D M   1988 Phys. Rev. Letts. $\bf 61$ 1029 ; Senthil T and Fisher M P A 2005 cond-mat/0510459 \\
$[3]$  Chaudhury Ranjan and  Paul Samir K 1999 Phys. Rev. B $\bf 60$  6234 \\
$[4]$  Chaudhury Ranjan and  Paul Samir K 2002 Mod. Phys. Letts. B $\bf 16$  251 \\
$[5]$  Chaudhury Ranjan and  Paul Samir K  ( To be communicated)\\
$[6]$  Loh E Jr.,  Scalapino D J and  Grant P M 1985 Phys. Rev. B  $\bf 31$  4712 ; Swendsen R H 1982 Phys. Rev. Lett.  $\bf 49$  1302 ; Betts D D , Salevsky F C and     Rogiers J  1981 J. Phys. A $\bf 14$ 531\\ 
$[7]$  Berciu Mona and  John Sajeev 2000 Phys. Rev. B  $\bf 61$ 16454 ; Mertens F G , Bishop A R, Wysin G M and Kawabata C 1987 Phys. Rev. Lett. $\bf 59$ 117 \\
$[8]$  Morinari T and  Magn J 2006 Mag. Mat. $\bf 302$  382 \\ 
$[9]$  M\'ol A S , Pereira A R , H. Chamati H and  Romano S 2006  Eur. Phys. J. B  $\bf 50$  541 \\
$[10]$ Cuccoli A ,  Valerio T , Paola V and  Ruggero V 1995 Phys. Rev. B $\bf 51$  12840 ; Kosterlitz J M and Thouless D J 1973 J. Phys. C $\bf 6$ 1181 ; Berezinskii  V  L 1970 Sov. Phys. JETP $\bf 32$ 493 , 1972 Sov. Phys. JETP $\bf 34$ 610      \\
$[11]$ Affleck Ian 1986 Nucl. Phys. B $\bf 265$ [FS15]  409 ; Ribeiro T C , Seidel A , Han J H and Lee D H 2006 Europhys. Lett. $\bf 76(5)$  891 \\

{\bf Appendix}

 Applying variational  principle  to the action given by equation (14), we  obtain the equation of motion  for an XY-anisotropic Heisenberg ferromagnet explicitly on the lattice , in the following form:
\begin{flushleft}
${{\partial_t}{ n_1}}(ia,ja)  =  -  i~g~s~[{\lambda_z}{ n_2}{N_3} - { n_3}{N_2}](ia,ja)$ 
\end{flushleft}
\begin{flushleft}
${{\partial_t}{ n_2}}(ia,ja)  =  -  i~g~s~[{ n_3}{N_1} - {\lambda_z}{ n_1}{N_3}](ia,ja)$ 
\end{flushleft}
\begin{flushleft}
${{\partial_t}{n_3}}(ia,ja)  =  -  i~g~s~[{ n_1}{N_2} -{ n_2}{N_1}](ia,ja)$\hfill(A.1)\\
\end{flushleft} 

In the above derivation we have used the following variation of the WZ-term given by (see Fradkin in Ref.[2])
\begin{flushleft}
$\delta {{\mathcal S}^{tot}_{WZ}}[\bf m]  =  {\sum_{i,j}}~\delta {{\mathcal S}_{WZ}}[{\bf m}(ia,ja)]$ 
\end{flushleft}
\begin{flushleft}
$=  {\sum_{i,j}}~{\int_0^{\beta}}~dt~\delta {\bf n}(ia,ja)\cdot ({\bf n}\wedge {\partial_t}{\bf n})(ia,ja)$\hfill(A.2)\\
\end{flushleft}
In Eqn.(A.1) ${N_1}(ia,ja),{N_2}(ia,ja),{N_3}(ia,ja)$ are the components of the vector $\bf N$ at the lattice point $(ia,ja)$ . The vector $\bf N$ is given as:
\begin{flushleft}
${\bf N}(ia,ja) = {\bf n}(ia,(j-1)a) + {\bf n}((i-1)a,ja) + {\bf n}((i+1)a,ja) + {\bf n}(ia,(j+1)a)$\hfill(A.3)\\
\end{flushleft}
 For the vortex ( anti-vortex ) plaquette having vertices 1 , 2 , 3 and 4 (see $Fig.1$ )   with coordinates $((i-1)a,(j-1)a)$, $(ia,(j-1)a)$, $(ia,ja)$ and $((i-1)a,ja)$ respectively , it  follows from Eqn.(10) that the contribution of  ${{\mathcal S}_{WZ}^{effective}}$  to the 1-vortex    is  given by 
\begin{flushleft}
$ {{[{{\mathcal S}_{WZ}^{effective}}]}_{(1-vortex)}}  =  - \{ {\int_0^{\beta}}dt [{\bf n}((i-1)a,ja)\cdot ({\bf n}\wedge {\partial_t}{\bf n}) (ia,ja)$
\end{flushleft}
\begin{flushleft}
$ +{\bf n}(ia,(j-1)a)\cdot ({\bf n}\wedge {\partial_t}{\bf n}) (ia,ja) + {\bf n}((i-1)a,(j-1)a)\cdot ({\bf n}\wedge {\partial_t}{\bf n}) (ia,(j-1)a)$
\end{flushleft}
\begin{flushleft}
$+{\bf n}((i-1)a,(j-1)a)\cdot ({\bf n}\wedge {\partial_t}{\bf n}) ((i-1)a,ja)] \}$\hfill(A.4)\\
\end{flushleft}
We evaluate the right hand side of $Eqn.(A.4)$ by substituting for  ${\partial_t}{\bf n}$ from  $Eqns.(A.1)~ and~ (A.3)$ . To keep the calculations  simple but consistent , we retain the intra-plaquette contributions by imposing a so called "local periodic boundary condition" (local PBC) as applied to the site closest to the vertices belonging to the plaquette under consideration .\\
In this case of 1-vortex [see Fig.1a] for the local PBC the field configurations satisfy
\begin{flushleft}
${\bf n}(ia,ja) = {\bf n}((i-2)a,ja) = {\bf n}(ia,(j-2)a)$
\end{flushleft}
\begin{flushleft}
${\bf n}((i-1)a,ja) = {\bf n}((i-1)a,(j-2)a) = {\bf n}((i+1)a,ja)$
\end{flushleft}
\begin{flushleft}
${\bf n}(ia,(j-1)a) = {\bf n}(ia,(j+1)a) = {\bf n}((i-2)a,(j-1)a)$
\end{flushleft}
\begin{flushleft}
${\bf n}((i-1)a,(j-1)a) = {\bf n}((i+1)a,(j-1)a) = {\bf n}((i-1)a,(j+1)a)$\hfill(A.5)\\
\end{flushleft}
 Besides we make use of  the following conditions satisfied  at different vertices of the plaquette for all time, as explained in the section 3.1
 [see  Fig.1a] :
\begin{flushleft}
${ n_1}((i-1)a,(j-1)a) = - { n_1}(ia,ja) = 1 - \delta $
\end{flushleft}
\begin{flushleft}
${ n_2}(ia,(j-1)a) = - { n_2}((i-1)a,ja) = 1 - \delta$\hfill(A.6)\\
\end{flushleft}
This means that the following equations must hold :
\begin{flushleft}
${\partial_t}{ n_1}((i-1)a,(j-1)a)  =  - {\partial_t}{ n_1}(ia,ja)  =  {\partial_t}(1 - \delta )$
\end{flushleft}
\begin{flushleft}
${\partial_t}{ n_2}(ia,(j-1)a)  =  - {\partial_t}{ n_2}((i-1)a,ja)  =  {\partial_t}(1 - \delta )$\hfill(A.7)\\
\end{flushleft}

 Using Eqns. (A.1) , (A.4) and (A.5) we have 
\begin{flushleft}
${{\partial_t}{ n_1}}((i-1)a,(j-1)a)  =  -i~g~s~{\lambda_z}{ n_2}{N_3}((i-1)a,(j-1)a) $
\end{flushleft}
\begin{flushleft}
${{\partial_t}{ n_1}}(ia,ja)  =  -  i~g~s~{\lambda_z}{ n_2}{N_3}(ia,ja)$
\end{flushleft}
\begin{flushleft}
${{\partial_t}{ n_2}}(ia,(j-1)a)  =  i~g~s~{\lambda_z}{ n_1}{N_3}(ia,(j-1)a)$
\end{flushleft}
\begin{flushleft}
${{\partial_t}{ n_2}}((i-1)a,ja)  =  i~g~s~{\lambda_z}{ n_1}{N_3}((i-1)a,ja)$ \hfill(A.8)\\               
\end{flushleft} 
 The right hand sides of Eqs. (A.8) vanish in the flattened meron configuration limit where ${\lambda_z}\longrightarrow 0$  and   ${N_3}$  aquiring very small since ${\epsilon}(ia,ja)$ is very small in this configuration. Consequently  from Eqns.(A.7) it follows that the quantity $ \delta $  does not vary with time   (  $\delta $  is independent of the position of the vertex points of the plaquette [See Fig.1.a] , as explained in section 3.1) \\
                                            
   Now using the Eqns. (A.1), (A.3),  (A.5), (A.6) and (A.7)  we obtain ~${{\mathcal S}_{WZ}^{effective}}$~   as  follows :
\begin{flushleft}
$ {{[{{\mathcal S}_{WZ}^{effective}}]}_{(1-vortex)}}   =  [{n_1}(2)+{n_1}(4)][({\bf n}\cdot {\bf N})(3) {n_1}(3) - {N_1}(3)]$
\end{flushleft}
\begin{flushleft}
$+[{n_3}(2)+{n_3}(4)][({\bf n}\cdot {\bf N})(3) {n_3}(3) - {N_3}(3)] + {n_1}(1)[({\bf n}\cdot {\bf N})(4){n_1}(4) + ({\bf n}\cdot {\bf N})(2){n_1}(2)]$
\end{flushleft}
\begin{flushleft}
$ + {n_2}(1)[({\bf n}\cdot {\bf N})(4){n_2}(4) + ({\bf n}\cdot {\bf N})(2){n_2}(2) - 2{ N_2}(2)] + {n_3}(1)[({\bf n}\cdot {\bf N})(4){n_3}(4) $
\end{flushleft}
\begin{flushleft}
$+ ({\bf n}\cdot {\bf N})(2){n_3}(2) - 2{ N_3}(2)]$
\end{flushleft}
\begin{flushleft}
$ =  A + B$\hfill(A.9)\\
\end{flushleft}

where $A$  and  $B$  are  given  by:
\begin{flushleft}
$A  = {\int_0^{\beta}}dt~~ \{  2{{({n_1}(2)+{n_1}(4))}^2}[{{n_1}^2}(3) - 1] + 2~{n_1}(3)~{n_3}(3)~[{n_1}(2)+{n_1}(4)][{n_3}(2)+{n_3}(4)]$
\end{flushleft}
\begin{flushleft}
$ + 4~{n_2}(1)~[{{n_2}^2}(2) - 1][{n_2}(3)+{n_2}(1)]   \} $
\end{flushleft}
\begin{flushleft}
$B  = {\int_0^{\beta}}dt~~ \{   + 2~{n_2}(2)~{n_1}(1)~[{n_1}(2)-{n_1}(4)]~[{n_2}(3)+{n_2}(1)]$
\end{flushleft}
\begin{flushleft}
$ + 2~{n_2}(2)~{n_3}(1)~[{n_3}(2)-{n_3}(4)]~[{n_2}(3)+{n_2}(1)] \}$\hfill(A.10)\\\end{flushleft}

It is interesting to note that $A$ remains invariant if we go from vortex to antivortex by changing  ${n_2}(2)$ and ${n_2}(4)$ in $Fig.1$ to $-{n_2}(2)$ and $-{n_2}(4)$ respectively whereas $B$ goes over to $-B$ . Thus  ~${{[{{\mathcal S}_{WZ}^{effective}}]}_{(1-vortex)}}$~  takes the form $A-B$ for antivortex $[Fig1.b]$.\\ 

To analyse the case of 2- vortex [Fig.2] we assign the co-ordinates to the vertices 1, 2, 3, 4, 5, 6, 7, 8 as $((i-1)a,(j-1)a)$ , $(ia,(j-1)a)$, $((i+1)a,(j-1)a)$, $((i+1)a,ja)$, $((i+1)a,(j+1)a)$, $(ia,(j+1)a)$, $((i-1)a,(j+1)a)$, $((i-1)a,ja)$ respectively .  The centre O  has co-ordinates  $(ia,ja)$. Using Eqns. (A.1) , (A.3) as well as the local PBC ; the equations motion  of the spins ${n_2}$ at the vertices  2, 4, 6, 8 reduce to the following , in the flattened  meron  configuration limit ( ${{\lambda}_z}\longrightarrow 0$ ):\\
\begin{flushleft}
${{(1-\delta )}^2} = {{cos}^2}{\epsilon (2)} = {{cos}^2}{\epsilon (4)} = {{cos}^2}{\epsilon (6)} = {{cos}^2}{\epsilon (8)}$\hfill(A.11)\\
\end{flushleft} 
It is therefore obvious from the above equations that the z-component of the spin $\bf n$ viz., $n_3$ becomes position indepedent.  \\
 In Fig.3 we have denoted the elementary vortex plaquettes by $\bf a$, $\bf b$, $\bf c$, $\bf d$, $\bf e$, $\bf f$, $\bf g$, $\bf h$. Note that there is an elementary anti-vortex plaquette in the central region with vertices $\bf A$, $\bf B$, $\bf C$ and $\bf D$ .  We write formally the expression for ~${{[{{\mathcal S}_{WZ}^{effective}}]}_{(3-vortex)}}$~ :\\
\begin{flushleft}
~${{[{{\mathcal S}_{WZ}^{effective}}]}_{(3-vortex)}}  =   {\int_0^{\beta}}dt   \{  {\bf n}(1a)\cdot ({\bf n}\wedge {\partial_t}{\bf n})(2a) +{\frac{1}{2}}{\bf n}(2a)\cdot ({\bf n}\wedge {\partial_t}{\bf n})(3a) +{\frac{1}{2}}{\bf n}(4a)\cdot ({\bf n}\wedge {\partial_t}{\bf n})(3a) + {\bf n}(1a)\cdot ({\bf n}\wedge {\partial_t}{\bf n})(4a)$
\end{flushleft}
\begin{flushleft}
$ + {\bf n}(1b)\cdot ({\bf n}\wedge {\partial_t}{\bf n})(2b) + {\bf n}(2b)\cdot ({\bf n}\wedge {\partial_t}{\bf n})(3b)+{\frac{1}{2}}{\bf n}(4b)\cdot ({\bf n}\wedge {\partial_t}{\bf n})(3b) +{\frac{1}{2}}{\bf n}(1b)\cdot ({\bf n}\wedge {\partial_t}{\bf n})(4b)$
\end{flushleft}
\begin{flushleft}
$ +{\frac{1}{2}}{\bf n}(1c)\cdot ({\bf n}\wedge {\partial_t}{\bf n})(2c) + {\bf n}(2c)\cdot ({\bf n}\wedge {\partial_t}{\bf n})(3c)+ {\bf n}(4c)\cdot ({\bf n}\wedge {\partial_t}{\bf n})(3c) +{\frac{1}{2}}{\bf n}(1c)\cdot ({\bf n}\wedge {\partial_t}{\bf n})(4c)$
\end{flushleft}
\begin{flushleft}
$+ ..............................$
\end{flushleft}
\begin{flushleft}
$+{\frac{1}{2}}{\bf n}(1h)\cdot ({\bf n}\wedge {\partial_t}{\bf n})(2h) +{\frac{1}{2}}{\bf n}(2h)\cdot ({\bf n}\wedge {\partial_t}{\bf n})(3h)+{\frac{1}{2}} {\bf n}(4h)\cdot ({\bf n}\wedge {\partial_t}{\bf n})(3h) + {\bf n}(1h)\cdot ({\bf n}\wedge {\partial_t}{\bf n})(4h)  \}$\hfill(A.12)\\
\end{flushleft}
In the above equation (1a), (2a), (3a) and (4a) denote the vertices for the subvortex $\bf a$ in the anti-clockwise sense as we have already usde in the case of 1-vortex. Similarly for the other subvortices. It can be shown after  a long calculation  that Eqn. (A.13)  can be written in the form  $A + B$ , as we had in the case of 1-vortex , where $B$  consists of terms  linear  in ${ n_2}(2)$ or containing odd powers of  ${ n_2}(2)$. Thus $A + B$ over goes to $A - B$ as we  go from vortex to anti-vortex.\\
In the case of a 4-votex [Fig.4] we have a again a single lattice point as in the case of the 2-vortex. By similar reasons the 4-vortex construction in terms of elementary plaquettes, breaks down in the flattened meron configuration limit.

\end{document}